\documentstyle[epsf,aps,prl,twocolumn,floats,eqsecnum]{revtex}

\begin{document}

\tighten
\bibliographystyle{prsty}

\title{
Geometrical phase effects in biaxial nanomagnetic particles
}

\author{Sahng-Kyoon Yoo\cite{e-ysk} }
\address{Department of Physics, Chongju University, Chongju, 360-764, Korea
         \\}
\author{Soo-Young Lee\cite{e-lsy} }
\address{Department of Physics, Kyungnam University, Masan, 631-701, Korea
         \\
\smallskip
{\rm (\today)}
\bigskip\\
\parbox{14.2 cm}
{\rm
The oscillation of tunnel splitting are obtained by
geometrical analysis of the topological Wess-Zumino phase on the basis of
tunneling paths in biaxial nanomagnetic particles with magnetic field along
the hard anisotropy axis. This theory not only just yields the previous
quantum interference results for the ground state tunneling,
but also gives the excited level splittings, all of which
agree well with the numerical diagonalization. Furthermore, the parity effect
in the asymmetric double well system which was recently discovered in
experiment, can be derived by similar arguments, and is also certified 
by using the complex periodic orbit theory.
The possibility of improving the discrepancy with the experiment 
in the periods of oscillations is discussed.
\smallskip
\begin{flushleft}
PACS number (s) :75.45.+j, 75.50.Tt, 03.65.Sq
\end{flushleft}
}
}

\maketitle

\section{INTRODUCTION}
\label{introduction}
For last decade, many physicists have explored the nanoscale magnetic
particles to study the extrapolation of quantum mechanics to macroscopic
realm. The magnetizations of these particles show the macroscopic quantum
tunneling (MQT) from a metastable state to a stable one, or macroscopic
quantum coherence(MQC) between two degenerate ground states separated by
potential energy barrier\cite{leggett87,gunther95,chudnovsky98}. 
In the latter case, the geometrical phase, known as the Wess-Zumino
phase\cite{fradkin91}, plays a crucial role in tunnel splitting. 
In particular, the biaxial nanomagnetic particles present some exotic
phenomena related to this topological phase. In the absence of external
magnetic field, the ground state tunneling vanishes only for half-integer
spins, which is called as spin parity effect and related to well-known
Kramer degeneracy\cite{loss92}. When the magnetic field
is applied along the hard anisotropy axis, however, the tunnel splitting
oscillates with magnetic field regardless of spin values\cite{garg93}. 
These spin parity effect and oscillations of tunnel splittings are
interpreted as the interferences of geometrical phases between two
opposite tunneling paths in the instantonic
approach\cite{loss92,garg93,gou99,chudnovsky99,yoo99}.

Most theoretical treatments so far about the oscillations of level splittings
have been restricted to the instanton method for the mapped effective
one-dimensional potential by performing the cos $\theta$ integration in spin
coherent state representation\cite{gou99,chudnovsky99} or by direct particle
mapping method\cite{yoo99}. In these approaches, although the ground state
tunneling can be obtained well, it is not possible to treat the tunneling
between excited energy levels exactly. In fact, the result of numerical
diagonalization for excited states is that the number of oscillations
decreases one by one as the quantum number increases.
In addition, due to the mapping onto one-dimension, the dominant tunneling
path between two wells in $(\theta, \phi)$ space is not clear. This makes
it difficult to perform the direct geometrical analysis of Wess-Zumino phase.
Recently, Garg\cite{garg99} derived the splitting oscillations in high energy
levels using the discrete WKB approximation. Another semiclassical approach
is the complex periodic orbit theory\cite{robbins89,robbins90}, the extension
of the trace formula in periodic orbit
theory\cite{gutzwiller67,balian70,berry76} to tunneling systems.
We obtained successfully the energy levels for not only the ground state
but also all the higher excited states by applying this method to spin system
for the first time, and showed the complete agreement with numerical
diagonalization\cite{lee00}. Since this theory also uses the effective mapping
potential obtained by cos $\theta$-integration of original Hamiltonian,
the dominant tunneling path in phase space remains to be unspecified yet.
The level splitting oscillations were confirmed by experiment very recently
in the Fe$_8$ molecular clusters with spin $S=10$\cite{wernsdorfer99}. 
They showed that, although the oscillations occurrs as
predicted by theory, in quantitative aspect the oscillation periods is about
1.6 times larger than previous theoretical results, and this discrepancy
can be resolved introducing the fourth-order terms in spin Hamiltonian.

In this paper, we find the dominant tunneling paths in phase space, and
analyze geometrically the Wess-Zumino phase without any mapping
in order to obtain the ground and excited level splitting oscillations
with magnetic field along the hard axis.
Furthermore, the experimentally discovered parity effect\cite{wernsdorfer99}
which is very similar to spin parity effect in the asymmetric system is derived
theoretically using this geometrical analysis and complex periodic orbit
theory. We also discuss the possibility of improving the discrepancy with
experiments by adding the fourth-order terms in spin Hamiltonian.
In section II, after finding the dominant tunneling path, we obtain
the splitting oscillations for the tunneling between both the ground and
the excited states in symmetric system
by investigating the relation between the geometrical
phase and the tunneling path in phase space without mapping onto a particle.
In Section III, the parity effect occuring in the asymmetric system is
derived using the arguement of Section II, and furthermore is also certified
from the complex periodic orbit theory. We discuss the possibility of
improving the quantitative discrepancy with experiment and give conclusions
in Section IV.

\section{LEVEL SPLITTING OSCILLATIONS IN THE SYMMETRIC SYSTEM}
\label{oscillations}

In this section, we calculate the both ground and excited level splitting
oscillations by analyzing the topological Wess-Zumino phase geometrically
in phase space. The tunneling rate in spin system is 
\begin{equation}
\Gamma = \int {\cal D} [\cos \theta] {\cal D} [\phi] \exp (-S_{E} )
\label{pathint}
\end{equation}
in the path integral formalism, where ${\cal D}$ means the integration over
all paths and $S_{E}$ is the Euclidean action which is given as
\begin{equation}
S_{E} = i S_{WZ} + \int d \tau {\cal H} (\theta, \phi).
\end{equation}
Here, $\tau$ is the imaginary time, ${\cal H} (\theta, \phi)$ is the spin
Hamiltonian in the $(\theta, \phi)$ representation, and
\begin{eqnarray}
S_{WZ} &  \equiv & S \int d \tau {\dot \phi}
\left[ 1 - \cos \theta (\tau) \right]
\nonumber \\
& \equiv & \int_0^{\pi} d \phi
\left\{ S - p [\phi(\tau)] \right\}
\label{wz}
\end{eqnarray}
is the Wess-Zumino action with total spin number $S$. Here
$p \equiv S \cos \theta$ and
corresponds to the conjugate momentum to coordinate $\phi$, since $\phi$ and
$p$ satisfy the Hamilton's equations of motion in the spin system.
When we consider the tunneling from $\phi=0$ to $\pi$, Eq. (\ref{wz}) means
the area surrounded by $p=S$, $\phi=0,~\pi$ lines and the tunneling path
$p(\phi)$ in $(\phi,p)$ phase space. [See the Fig. \ref{figure1} (a).]
Since the energy splitting due to tunneling, $\Delta E$, is determined
by Wess-Zumino action (the area mentioned above), i.e.,
\begin{equation}
\Delta E \sim \cos (S_{WZ}),
\label{dele}
\end{equation}
whenever this area becomes $(n + 1/2) \pi$ with integer $n$ the tunnel
splitting vanishes. In order to perform the geometrical evaluation of
Wess-Zumino phase without mapping, it is essential to find the dominant
tunneling path $p(\phi)$ in the phase space.
It will be shown below that in the case of $k$th state,
the Wess-Zumino phase continuously decreases from $(S-k)\pi$
to zero as the external field $h$ increases, so that the moment of area
being the half-integral multiple of $\pi$, at which the tunnel splitting
quenches, occurs $(S-k)$ times.

Let us consider the Hamiltonian describing the biaxial spin system which
is given by
\begin{equation}
{\cal H}= -D S_x^2 + E(S_z^2 - S_y^2 )+g \mu_{B} {\bf H} \cdot {\bf S},
\label{ham0}
\end{equation}
where $D$ and $E$ are longitudinal and transverse anisotropy constants,
respectively, $g$ is the gyromagnetic ratio and $\mu_{B}$ is the Bohr
magneton. This Hamiltonian represents that the system has an easy axis
in the $x$-direction, and hard axis in the $z$-direction.
When the magnetic field ${\bf H}$ is applied along the $z$-direction,
the system is symmetric and
the reduced Hamiltonian is given in the spin coherent state representation by
\begin{eqnarray}
{\cal H'} & = & [ 2(1-\lambda) + (2 \lambda -1) \cos^2 \phi ] p^2 - 2 S h p 
\nonumber \\
& & - (1- \lambda) S^2 - (2 \lambda -1) S^2 \cos^2 \phi,
\label{ham1}
\end{eqnarray}
where ${\cal H'} \equiv {\cal H} /(D+E)$, $\lambda=D/(D+E)$ and 
$h \equiv H/H_c =g \mu_{B} H/2 (D+E) S$ with coersive field $H_c$. 
In $(p, \phi)$ phase space, ${\cal H'}$ has two degenerate minima
when $p=S h$ and $\phi=0, \pi$. At $\phi= \frac{\pi}{2}$, however,
$p= Sh/ h_1$ with $h_1 \equiv 2 (1- \lambda)$ gives the saddle point
for $0<h< h_1$, local minimum for $h_1<h< h_2$ with
$h_2 \equiv \sqrt{2 (1- \lambda)}$ and global minimum for $h_2 <h<1$,
respectively, if $p$ can be also defined beyond the range $-S \le p \le S$.
These pictures of various external fields are shown in Fig. \ref{figure1}
as the energy contour plots.

In order to get an information about the dominant tunneling path, we revisit,
for a moment, the mapping onto a one-dimensional particle problem by integration
over $\cos \theta$ in Eq. (\ref{pathint}). The resulting Euclidean action
is given by
\begin{equation}
S_E (\phi) = i A(\phi) + S \sqrt{2 \lambda-1} \int d \tau
\left[ \frac{1}{2} M(\phi) {\dot \phi}^2 + V(\phi) \right],
\label{euclact}
\end{equation}
where the dot means the Euclidean time derivative and the imaginary part of
Euclidean action which is responsible for the phase effect is
\begin{equation}
A (\phi)= S \int d \phi \left[ 1- \frac{h}{1-(2 \lambda-1) \sin^2 \phi} \right],
\label{wzphase}
\end{equation}
the $\phi$-dependent effective mass is
\begin{equation}
M (\phi) = \frac{1}{1- (2 \lambda-1) \sin^2 \phi}
\end{equation}
and the effective potential is
\begin{equation}
V(\phi)= \frac{1}{2} \sin^2 \phi \left[ 1- \frac{h^2}{1- (2 \lambda-1) \sin^2 \phi}
\right].
\label{effectivev}
\end{equation}
For the ground state tunneling, if we compare Eq. (\ref{wz}) and
Eq. (\ref{wzphase}), the tunneling path $p(\phi)$ is found to be the second

It is emphasized that the tunneling path $p(\phi)$ in the phase space
$(\phi,p)$ is just the path satisfying $\partial H'(\phi, p)/\partial p=0$.
This means that the tunneling path follows the minimum energy positions at
all $\phi$ values. These tunneling paths for vaious fields are shown in
Fig. 1 as thick solid lines. This simple condition for dominant tunneling
path in the phase space $(\phi, p)$ works well only when the Gaussian
integration is possible as in the present case.

Now, we calculate the Wess-Zumino action (the area) mentioned above as $h$
increases from 0 to 1. When $h=0$, the phase is just $S \pi$
(whole area of phase space considered) because the dominant tunneling path
is $p(\phi)=0$. It is important to note that, although the quantum
mechanical ground state energy is not the well minimum, the path starts
and ends at well minima, because the Euclidean action with the energy of
the minimum gives the information about the ground state splitting in
semiclassical theory. When $0< h < h_1$ [Fig. \ref{figure1} (a)],
the path is confined within a range $0<p<S$. Therefore, we can easily
calculate the area enclosed by the path and $p=S$ line.
Until $h$ becomes $h_1$, where the north-pole becomes the saddle point
of energy, the area continuously decreases from $S \pi$.
For $h_1 < h < h_2$ [Fig. \ref{figure1} (b)], however, the local minimum
exists outside the $p=S$ line. In this case the tunneling path runs beyond
the $p=S$ line and thus, the Wess-Zumino action becomes the area of
(region B + region C - region A). As $h$ increases the action continues
to decrease and becomes zero at $h=h_2$. As a result, within a range
$0<h<h_2$, the quenching take places $S$ times for the ground state
by Eq.(\ref{dele}). The range $h_2 < h <1$ is more or less subtle
[Fig. \ref{figure1} (c)]. The local minimum turns into the global one,
i.e., the point $(Sh/h_1, \pi/2)$ becomes lower than the well minimum.
In this case, the path giving the Wess-Zumino action should be determined
by somewhat different way, because there exists the contour of $E_0$
(well minimum energy). If the path obtained by the previous method
encounters the contour of $E_0$, it should follow the contour in a real time
due to energy conservation. Therefore, the action is equivalent to the area
of [region B + region C - region A ] of Fig. 1(c) which always vanishes
in the range of $h_2 < h < 1$.
This vanishing of the action is closely related to the cancellation between
resonant tunneling and quenching phenomena\cite{lee00} and also between
the Euclidean action of real time motion and Wess-Zumino
action\cite{chudnovsky99}.

The extention to the excited states is straightforward.
The $k$th energy level of each well can be calculated from the well-known
EBK (Einstein-Brillouin-Keller) quantization rule\cite{einstein17},
\begin{equation}
\oint p (\phi) d \phi = 2 \pi \left( k + \frac{1}{2} \right),
~~~k=0,1,2, \cdots,
\end{equation}
In this quantization, whenever the action for closed orbit in one well differs
by $2 \pi$, the energy level is given. It will be shown that, by this fact,
the number of splitting oscillations decreases one by one with increasing
quantum number. When $h=0$, there exist the contours of $E_k$ centered at
$p=0$ and $\phi=0, ~\pi$ whose areas are $2 k \pi$ for $k$th excited state
(see also Fig. \ref{figure1}). If we use the analogy for the case
$h_2 < h < 1$ of the ground state tunneling, the area of $k \pi$ is
excluded, due to the energy conservation, for the $k$th excited state
($k \pi/2$ for each well) in the phase space considered,
i.e., $(S-k) \pi$. As $h$ increases the continuous reduction to zero is
same as the ground state case. This leads to the $(S-k)$ oscillations
for $k$th energy level. The calculated periods of oscillations for both
ground and excited state levels completely agree with the numerical
diagonalizations. Comparing with the previous semiclassical
theories\cite{garg93,gou99,chudnovsky99,yoo99},
in this theory the dominant tunneling path in $(\phi, p)$ phase space
can be found to make possible to perform the geometrical analysis of the
effect of topological phase, and in addition to the gound state case
the structures of splitting oscillations for the excited states can be
obtained easily on the basis of EBK quantization scheme.

\section{PARITY EFFECT IN THE ASYMMETRIC SYSTEM}
\label{parity}

Unlike the symmetric case in Sec. II, in this section we treat the
asymmetric system having the longitudinal field as well as the
transverse field. In such system, we derive the parity effect 
which was discovered in recent experiment\cite{wernsdorfer99}, by using
the simple geometrical argument developed in section II, and also lead to
the same result in the semiclassical complex periodic orbit theory.
This parity effect is very similar to the spin parity effect proposed
by Loss et. al. and von Delft and Henley\cite{loss92}. 
This is observed when one measures the transtion between the
ground state in upper well and the $k$th excited  state in lower well,
by applying the external field $h'$ along the easy axis (longitudinal field)
which makes the system asymmetric in addtion to $h$ in the hard direction
(tansverse field). For the case of the tunneling from the ground state
of the upper well to even $k$th state of lower well, the phase of the
oscillation is same as that of the symmetric case, while for the case of
tunneling to odd $k$th state of the lower well the phase is shifted by
$\pi /2$.

This interesting observation can naturally arises in our above analysis.
As shown in Fig. 2, the energy structure of spin system is asymmetric
due to $h'$. At certain values of $h'$ the ground state energy level $E_0^l$
in upper (left) well exactly coincides with the $k$th excited level $E_k^r$
in lower (right) well, so that one can consider the tunnel splittings.
The tunneling path starts from the minimum at upper well, and when it exits
the barrier it follows the contour of $k$th excited state in the lower well
due to energy conservation. Therefore, by EBK quantization relation and
analogous arguments with the excited states in Section II, the area
$k \pi/2$ in the $(\phi, p)$ phase space is excluded when $h=0$ case. 
As the transverse field $h$ increases, total area surrounded by this path
continuously decreases from $(S \pi - k \pi /2)$ to zero in the same manner
with the symmetric case except the $k \pi/2$ shift. This leads to the
parity effect in the asymmetric system.

Next, let us consider this effect in complex periodic orbit theory.
In quantum mechanics the energy spectrum of the system can be derived from
the singularities of the trace of the energy-dependent Green's function
which is given by
\begin{equation}
g(E)= {\rm Tr} \left[ (E - {\hat H})^{-1} \right] = \sum_{n} \frac{d_n}{E-E_n},
\label{dosq}
\end{equation}
where Tr means the trace and $d_n$ is the degeneracy factor.
The semiclassical approximation of Eq. (\ref{dosq}) was given by 
Gutzwiller\cite{gutzwiller67} in the form
\begin{equation}
g_{{\rm sc}} (E) = \frac{1}{i} \sum_{j} T_j \sum_{r=1}^{\infty}
e^{i r (S_j - \mu_j \pi/2)},
\label{dosc}
\end{equation}
where $T_j$, $S_j$ and $\mu_j$ are the period, the action and the Maslov index
of the $j$th primitive classical periodic orbit, and index $r$ corresponds
to the repetition of primitive periodic orbit. 
In evaluating this sum, we restrict on the one-dimensional effective
particle problem. The system with $h'=0$ which are given as
Eq. (\ref{euclact}) - Eq. (\ref{effectivev}) is recently treated in detail
within the complex periodic orbit theory in Ref. \cite{lee00}.
When the longitudinal magnetic field $h'$ is applied, the effective
potential of Eq. (\ref{effectivev}) becomes asymmetric as shown in Fig. 3.
This figure presents the case when the ground state energy of upper well
coincides with the $k$th excited level of lower well. The classical orbits
in two wells and tunneling orbits within the barrier are represented
as solid and dotted lines, respectively.
The directions denoted by arrows are given by the rules of
Ref. \cite{robbins89}.

Now we should consider all possible periodic orbits in order to perform
the summation in Eq. (\ref{dosc}). Let us take the classical segments
in the respective wells as the half of whole orbit having the actions
$W_1 \equiv S_1/ 2$ and $W_2 \equiv S_2/2$, and similarly
for the tunneling segments within the barrier having an action
$\Theta \equiv S_c /2$. Let us consider the case where the ground state
energy in the upper well is equal to the $k$th excited state in the lower
well. Note, then, that the relation between two actions $W_1$ and $W_2$ is
\begin{equation}
W_2 = W_1 + k \pi ,~~~k=0,1,2, \cdots,
\end{equation}
due to EBK quantization rule. The possible periodic orbits can be
parametrized as follows: i) the paths that start and end at
$\phi_{A}$ or $\phi_{A'}$ without complete rotations,
ii) the paths that start at $\phi_{A} (\phi_{A'})$ and end
at $\phi_{A'} (\phi_{A})$, i.e., having complete rotations
(See Fig. 3. $\phi_A$ and $\phi_{A'}$ are the equivalent position
in $\phi$ coordinate). The Wess-Zumino phase contributes only
to the paths belonging to category ii) as
\begin{equation}
\alpha = 2 \pi S \left(1- \frac{h}{h_2} \right).
\end{equation}
For the paths that start at $\phi_{A}$ and end at $\phi_{A'}$, i.e.,
proceed from left to right, we take the Wess-Zumino phase as positive,
whereas for the paths from right ($\phi_{A'}$) to left ($\phi_{A}$)
we take as negative. The Maslov index is easily evaluated
following the method for double well potential in Ref. \cite{robbins89},
i.e., each classical segment $W_1$, $W_2$ accompanies the additional phase
$+\pi/2$, while the tunneling segment $-\pi/2$. The all possible primitive
periodic orbits are drawn graphically in Fig. \ref{figure4},
where the solid line corresponds to $\exp[i (W_1+ \pi/2)]$,
thick solid line to $\exp[i (W_2 + \pi/2)]$ and dotted line to
$\exp[- \Theta -i \pi/2]$. Therefore, the semiclassical trace of
Green's function can be expressed as the summation over all repetitions
of these orbits, and its poles are found to be when
\begin{eqnarray}
& 1-e^{i(2 W_1+\pi)}+e^{-2 \Theta} +
\frac{e^{-2 \Theta}}{1-e^{i(2 W_2+\pi)}+e^{-2 \Theta}} &
\nonumber \\
& \times \bigg[ e^{i(W_1 +W_2 +\pi)}
\left( e^{i \alpha} + e^{- i \alpha} \right) &
\nonumber \\
& e^{i (2W_1 +\pi)}+e^{i (2 W_2+ \pi)} \bigg] = 0 &
\label{poles}
\end{eqnarray}
If we neglect, for the moment, the tunneling probability $e^{-2 \Theta}$, 
then the poles give the EBK energies $E_n$ for upper well where
\begin{equation}
S_1 (E_n) = 2W_1 (E_n) = 2\pi \left( n+ \frac{1}{2} \right),
\end{equation}
with $n=0$ for the ground state. Therefore, we can expand the actions
around the EBK energy like
\begin{eqnarray}
2W_1 (E) & = & \pi + T_1 (E- E_0 ) + \cdots ,
\nonumber \\
2W_2 (E) & = & \pi + 2k \pi + T_2 (E- E_0 ) + \cdots
\end{eqnarray}
near the ground state of upper well, where $T_1$ and $T_2$ are given by
$dS_1 /dE$ and $dS_2 /dE$, respectively.
Here, we used the fact that the difference of two actions at EBK energy
is $2 \pi k$ for the transition between ground state in upper well
and $k$th excited state in lower well. Expanding Eq. (\ref{poles})
up to the first-order in $e^{- 2 \Theta}$,
\begin{equation}
T_1 T_2 (E - E_0 )^2 - e^{-2 \Theta} \left[ (-1)^k
\left( e^{i \alpha} + e^{-i \alpha} \right) +2 \right] =0
\end{equation}
The final energy splitting is obtained as
\begin{eqnarray}
E & = & E_0 \pm \frac{2}{\sqrt{T_1 T_2}} e^{- \Theta} \cos \frac{\alpha}{2},
~~~{\rm even}~ k, \nonumber \\
E & = & E_0 \pm \frac{2}{\sqrt{T_1 T_2}} e^{- \Theta} \sin \frac{\alpha}{2},
~~~{\rm odd}~ k,
\end{eqnarray}
which is just the parity effect in the asymmetric system of
Ref. \cite{wernsdorfer99}.

\section{DISCUSSIONS AND CONCLUSIONS}
\label{conclusion}

Until now, we have analyzed the Wess-Zumino phase of biaxial spin system
in $(p, \phi)$ phase space, and obtained not only the same results with
those in the mapping formalism for ground state, but also
the excited energy level splitting oscillations. Furthermore, 
the parity effect in the asymmetric system shown in recent
experiment\cite{wernsdorfer99} is derived by both the geometrical analysis
and the complex periodic orbit theory.
However, the quantitative discrepancy with experiment in the period
of oscillation exists, i.e., about 1.6 times larger oscillation period
than expected in the theory. In order to resolve these discrepancies
the authors of Ref. \cite{wernsdorfer99} introduced the fourth-order term
$C (S_+^4 + S_-^4)$ in spin Hamiltonian (Eq. (\ref{ham1})), where $C$ is the
adjustable parameter which is shown to be $-2.9 \times 10^{-5} $K
in Kelvin unit through numerical diagonalization, when anisotropy constants
are given by $D=0.292 $K and $E=0.046 $K. Then the fourth-order Hamiltonian
is expressed in the $S_x$-representation as
\begin{equation}
{\cal H}_1 = {\cal H}' + 2C ( S_z^4 + S_y^4 - 6 S_z^2 S_y^2 )
\label{ham4}
\end{equation}
within the semiclassical approximation. In fact, if we consider the classical 
commutation relation satisfied by angular momenta, the cubic and quadratic
terms must be added in this Hamiltonian. But we have ignored them since
their contributions are negligible. It is not possible to get an anaytical
effective potential through integration, since Eq. (\ref{ham4}) is beyond
the Gaussian approximation. However, the qualitative analysis of
this Hamiltonian can be performed within our theory on the basis of
the energy structure of ${\cal H}_1$ and the dominant tunneling paths
in $(\phi, p)$ phase space.

The structure of energy barrier for ${\cal H}_1$ is different from that
for ${\cal H}'$ as shown in Fig. \ref{figure5}. Since $C$ is negative,
the effect of $S_z^2 S_y^2$ in Eq. (\ref{ham4}) is to raise the barrier
around $\phi= \pi/2$ and the mid-value of $0<p<S$, whereas $S_z^4
(S_y^4 )$ to lower the barrier at $z(y)$-axis.
Therefore, as the field $h$ increases, the saddle point moves more slowly
to the north-pole, compared with the $C=0$ case. If we use the same analogy
about finding the dominant tunneling path, this means that the slower decrease
of the area surrounded by tunneling path, and thus the larger oscillation
period. Although it is possible to get a qualitative correction to $C=0$
case in the desirable direction, in order to be consistent with experiment,
the adjustable parameter $C$ must be about 3 times greater than the present
value. The quantitative disagreement implies that the tunneling paths
which follow all energy minima are no longer the dominant paths
in the quartic case. In order to treat the quartic case exactly,
new theoretical approach in order to obtain the dominant tunneling path
in phase space is needed.

In conclusion, we derive both the ground and excited tunnel splitting
oscillations in the biaxial nanomagnetic particle with the magnetic field
along the hard anisotropy axis by finding the dominant tunneling path and
geometrically analyzing the topological Wess-Zumino phase
in the phase space. All the results of analysis are in agreement with the
numerical diagonalization. Furthermore, the interesting parity effect 
in the asymmetric case is naturally clear in this analysis, and also
certified within the complex periodic orbit theory. We also discussed
the possibility of improving the discrepancies with experiment in the period,
by introducing the quartic terms in spin variables into Hamiltonian.
In order to resolve the discrepancy quantitatively, the new approach to
find the dominant tunneling path in quartic case is required.
\\ \\

ACKNOWLEDGMENTS \\ \\
We thank to Dr. Cheol-Hong Kim for very helpful discussions on mathematics.

\newpage

\begin{figure}
\caption{ The contour plots of Hamiltonin Eq. (\protect\ref{ham1}) 
and the tunneling paths (thick solid lines) which follow all the minima with $p$ for
various external magnetic fields $h$. Here, $\lambda =0.861$ of Fe$_8$ is used.
In this case, $h_1=0.278$ and $h_2=0.528$.
The dashed lines mean $p=S$ lines.
a) $h=0.2$. The area corresponding to the geometrical phase is the region between
the tunneling path and $p=S$ line.  (b) $h=0.5$. The phase is evaluated as the
area of (rigion B + region C - region A). (c) $h=0.7$. When the path encounters the
contour of $E_0$, it follows the contour in real time (thick solid line with arrow).
The phase is also the area of (region B + region C - region A) but yields zero.
For the excited states, see the text.
}
\label{figure1}
\end{figure}

\begin{figure}
\caption{The typical asymmetric contour plot of the Hamiltonian with both 
longitudinal ($h'$) and transverse ($h$) magnetic fields.
The dominant path starts at the left well minimum $E_0^l$, and when it encounters
the contour of $k$th excited state in right well $E_k^r (=E_0^l)$, it follows the contour
(thick solid line with arrows).
}
\label{figure2}
\end{figure}

\begin{figure}
\caption{The asymmetric effective potential when the longitudinal
field is applied. The actions are denoted by $S_1$, $S_2$ and $S_c$.
Actually, $\phi_A$ and $\phi_{A'}$ are same positions, but we use the different
symbols in order to distinguish the direction.
}
\label{figure3}
\end{figure}

\begin{figure}
\caption{The diagram representation of all possible periodic orbits
in the asymmetric potential. The segments represent $e^{i(W_1 + \pi/2)}$
(solid line),  $e^{i(W_2 + \pi/2)}$ (thick solid line) and  $e^{- \Theta - i \pi/2}$
(dotted line).
}
\label{figure4}
\end{figure}

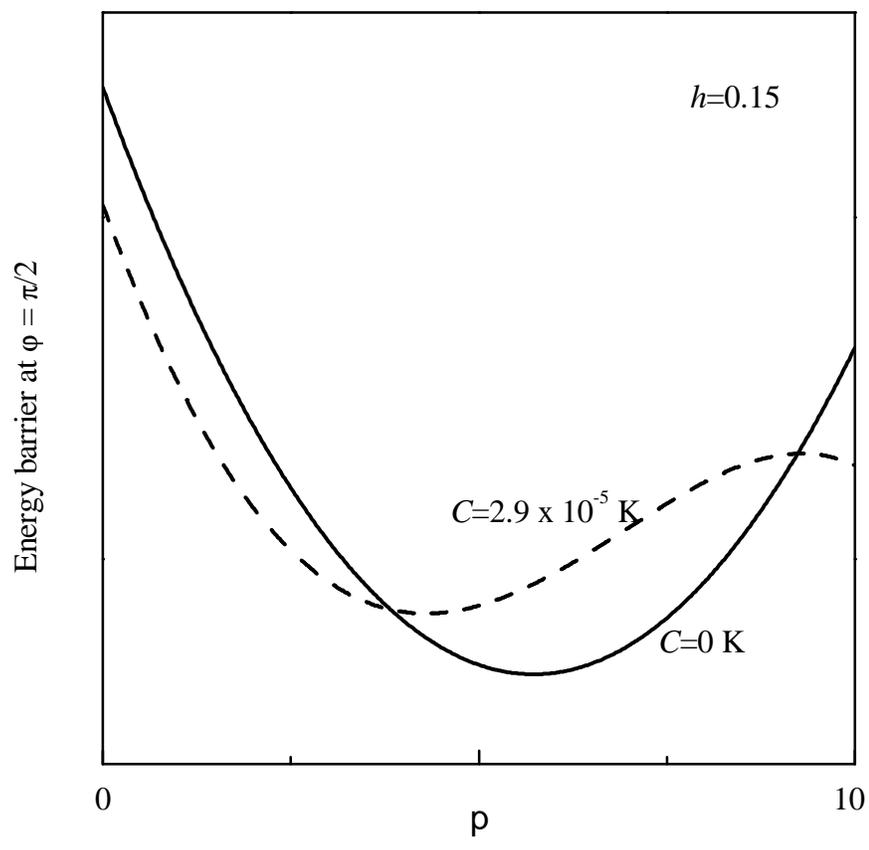
\begin{figure}
\caption{The energy barrier of Hamiltonians ${\cal H}_1 (p, \phi= \pi /2)$
for $C=0$ K (solid line), and $C=-2.9 \times 10^{-5}$ K (dashed line) at $h=0.15$.
As $h$ increases, the barrier minimum (saddle point) moves to the right. 
But the movement of that for $C=0$ K case is faster than for $C= 2.9 \times
10^{-5}$ K case. If we can think these minima as the passage that is taken by
the dominant tunneling path, the area reduction in phase space is smaller for
$C=2.9 \times 10^{-5}$ K case.
}
\label{figure5}
\end{figure}

\end{document}